\documentstyle[12pt,epsf,ssi]{article}

\begin{document}
\begin{flushright}
SLAC--PUB--7361\\
November 1996
\end{flushright}
\bigskip\bigskip

\thispagestyle{empty}
\flushbottom
\vfill
\centerline{{\large\bf QCD: Questions, Challenges, and Dilemmas}
    \footnote{\baselineskip=14pt
     Work supported by the Department of Energy, contract 
     DE--AC03--76SF00515.}}
\vspace{22pt}

  \centerline{\bf James D. Bjorken}
\vspace{8pt}
  \centerline{\it Stanford Linear Accelerator Center}
  \centerline{\it Stanford University, Stanford, California 94309}
\vspace*{0.9cm}

\begin{center}
Abstract
\end{center}
An introduction to some outstanding issues in QCD is presented, with
emphasis on work by Diakonov and co-workers on the influence of
the instanton vacuum on low-energy QCD observables.  This includes the
calculation of input valence-parton distributions for deep-inelastic
scattering.

\vfill

\centerline{Introductory talk given at the}
\centerline{XXIV SLAC Summer Institute on Particle Physics}
\centerline{``The Strong Interaction from Hadrons to Partons''}
\centerline{SLAC, Stanford, California}
\centerline{19--30 August 1996}
\vfill

\newpage

\renewcommand{\bar}[1]{\overline{#1}}
\newcommand{\etal}{{\em et al.}}
\newcommand{\ie}{{\it i.e.}}
\newcommand{\eg}{{\it e.g.}}
\newcommand{\D}{{\cal D}}
\newcommand{\nablaslash}{\not{\hbox{\kern-3pt $\nabla$}}}
\newcommand{\Aslash}{\not{\hbox{\kern-3pt $A$}}}
\newcommand{\Dslash}{\not{\hbox{\kern-4pt $D$}}}
\newcommand{\calDslash}{\not{\hbox{\kern-4pt $\D$}}}
\def\VEV#1{\left\langle{ #1} \right\rangle}
\hyphenation{quarks BFKL}

\section{Introduction}

{\it 
Quantum electrodynamics (QED) is the correct theory of
electromagnetism.}
\medskip

\noindent
{\it Quantum Chromodynamics (QCD) is the correct theory of the strong
force.}
\medskip

These bold, bald statements are slightly unscientific. Nevertheless
they are not far from the truth, in the sense that to challenge either
is a very serious enterprise, and one that is likely not to bear fruit
unless the challenge is an especially incisive one.

It is remarkable to me that in the short span of two decades QCD has
attained a degree of credibility competitive with QED: the truth and
the degree of falsity of the lead sentences above are at a comparable
level for the two theories. In fact we know that QED at short distances
does break down. The noble photon becomes the offspring of an ugly,
unaesthetic $U(1)$ gauge boson and the neutral $SU(2)$ electroweak
boson. Nothing like that fate appears to await the gluons, at least
this side of the GUT scale.

Both QED and QCD live in the family of gauge theories and are
structurally similar. Their Lagrangian densities both are $E^2-B^2$.
Both require the gauge-invariant substitution $p\rightarrow p - eA$. 
The Heavy-Quark Effective Theory of QCD has its counterpart in the
Heavy-Nucleon Effective Theory of QED, responsible for the
nonrelativistic limit of electrodynamics, which contains the
foundations of condensed matter theory, chemistry, biology, and more.

Both QED and QCD have their Feynman-diagram perturbation-theory
processes, leading to incisive precision tests---which work. Their
coupling constants run and are seen to run. QED and QCD are very well
``tested".

But just as nonperturbative QED contains very interesting phenomena, as
mentioned above, nonperturbative QCD is a most interesting portion of
that theory as well. To me it is {\it the} most interesting and most
important portion of QCD to address, despite the evident difficulty in
doing so. The lectures in this school emphasize the doable,
perturbation-theory based piece of QCD, because that is where most of
the work is occurring. In this introduction I have decided to try to
highlight the opposite extreme, with emphasis on material not covered
in the other lectures, as well as on the troubles, not successes. I
will omit some other unconventional QCD topics which I regard as
especially relevant to future high-energy collider experimentation,
because they are covered in another talk given to the Snowmass workshop
earlier this summer.\cite{ref:a}

\section{Questions}

These are rather random, just to set the tone. First some easy ones:

{\bf Q1.}\ 
Does the force between quarks get weaker at short distances?

{\bf A1.}\ 
You had better answer no. The force follows an approximate
inverse-square law, with a coefficient which at short distances very
slowly gets smaller (asymptotic freedom). Please don't accuse me of
nit-picking. It may be acceptable for us to use sloppy language to each
other but it is definitely very wrong when trying to explain QCD to the
outside world at the Scientific American level.

Say it right! When you do, it becomes perfectly clear why there are so
many high-$p_t$ jets in hadron-hadron collisions, jets that justify the
livelihood of so many experimentalists and theorists. The forces
between quarks get so incredibly strong that 500 $GeV$ partons which
collide head-on can make the right-angle turn at rates high enough to
be detected.

{\bf Q2:}\ 
In idealized QCD, with light quarks omitted, does the force between
quarks grow as their separation becomes very large?

{\bf A2:}\  
Again, no. It's the potential energy that grows linearly. 

{\bf Q3:}\ 
Is the QCD strong force $CP$-conserving?

{\bf A3:}\ 
In general, no. There is a $CP$-violating term {\it E.B} in the
Lagrangian which is allowed and admits observable effects like a
nonvanishing neutron electric dipole moment. Renormalization effects
make the coefficient of the $CP$-violating term formally divergent, but
the actual coefficient is very small, less than $10^{-9}$. What to do
remains an unsolved problem, probably not mentioned again in this
school.

{\bf Q4:}\ 
Do instantons matter?

{\bf A4:}\ 
Yes. These will not appear in other lectures but will be mentioned
in this one later on. They impact on, among other things, the $CP$
violation issue mentioned above.

{\bf Q5:}\ 
Does old fashioned pre-QCD $S$-matrix theory have anything to do with QCD?

{\bf A5:}\ 
Yes. While there seems to be a feeling that quarks, QCD and parton
ideology have rendered that body of work obsolete, this is not true.
The $S$-matrix techniques were built from general principles
(analyticity, unitarity, microscopic causality, crossing symmetry,
spectrum, $\cdots$) which are rigorously true in QCD.\cite{ref:c} 
Much can still be salvaged from these ideas in describing the
nonperturbative, confining, low-energy limit of QCD. It is still
something worth learning, and I fear that it is taught less and less,
much being eventually lost and having to be someday rediscovered afresh.

{\bf Q6:}\ 
Does Regge-Pole theory have anything to do with QCD?

{\bf A6:}\ 
This question is a special case of the previous one, with the same
answer, but with very clear implications, for example, in main-line QCD
structure-function phenomenology. Nonsinglet deep-inelastic structure
functions in the limit of small $x$ should be describable by exchange
of well-established Regge-trajectories like the $\rho$ or $\omega$.
These Reggeons are very well-established experimentally and precisely
parametrized. There is much less uncertainty in the theoretical
underpinnings of the asymptotic limit of nonsinglet structure functions
than there is in the related world of soft and hard Pomeron physics, to
be described by Al Mueller in this school.\cite{ref:d} Nevertheless
there is very little work going on to understand this problem in the
context of QCD, perturbative or otherwise.\cite{ref:e} It is becoming
of special current interest because of the experimental situation
regarding the small-$x$ behavior of the polarized structure
functions.\cite{ref:f}

{\bf Q7:}\ 
Is the boundary between what is legally calculable from
perturbation theory and what is not well defined?

{\bf A7:}\ 
I believe not. Furthermore it seems to be crossed more and more
indiscriminately as time goes on. Many calculations treat initial and
final quarks and gluons as on-shell, asymptotic states. This is
illegal; there is no $S$-matrix for quark and gluon interactions. At a
less fundamental level, some perturbative-QCD-inspired models for
hadronization push shamelessly into regions of parameter space (small
momenta, large distance scales) which are indefensible. While boldness
in this regard is in itself no vice, an uncritical attitude is. It is
not enough to say ``It agrees with data, therefore it makes sense and
is a prediction of the perturbative theory."

{\bf Q8:}\ 
Will these questions ever end?

{\bf A8:}\ 
Yes, right now.

\section{Challenges}

The basic challenges in understanding QCD can be seen very clearly in a
space-time description: it is how to link the phenomena at short
distances with phenomena at large distances. The simplest case is the
static limit, with all light quark degrees of freedom left out. The
short-distance limit is that of onium physics---a Coulomb-like
interaction between heavy quarks with a weak coupling constant. This is
under very good theoretical control. As the heavy quarks are pulled
apart there emerges a linear potential between them, something
described  quite well via the lattice calculations.\cite{ref:g} The
microscopic picture is believed to be that there is a color-electric
flux tube of smallish diameter between quark and antiquark in this
limit. However the dynamics creating it is the essence of the problem
of confinement and not ``understood" well. And if light quarks are
included, long flux tubes invariably break and are terminated by
constituent quarks or antiquarks. Pull apart bottomonium and you get a
$B$-meson and anti-$B$-meson. A $B$-meson is (by definition!) a
constituent quark plus a heavy spectator $b$-quark which can be treated
perturbatively. Therefore the $B$-meson dynamics is an especially
simple way in principle (alas, not so much in experimental practice) of
learning about the properties of single, ``isolated", constituent
quarks.\cite{ref:h}

Challenges for ``pure" QCD with light quarks excluded include the
understanding of the glueball spectrum,\cite{ref:i} as well as the
details of the flux tube. When the light quarks are introduced, there
are major changes to deal with: the glueballs mix with the myriad of
ordinary meson excitations of $q-\bar q$ pairs, perhaps toward the
limit of total extinction. Flux tubes break, but the microscopic
description is obscure.  Perhaps the flux-tube concept is likewise
driven to the edge of extinction.

Another very basic challenge for the static picture is the nature of
chiral symmetry breaking. Because the bare masses of up and down quarks
are so small, the QCD Lagrangian has an almost exact $SU(2)_L \times
SU(2)_R = O(4)$ chiral symmetry.  These $SU(2)$'s describe
independent isospin rotations of left- and right-handed up and down
quarks. There is a vacuum condensate $\langle 0|\sigma|0\rangle
\neq 0$, with $\sigma$ the fourth component of an internal-symmetry
four-vector ($\sigma, \vec\pi$) built from the quark densities. The
situation is very analogous to the Higgs sector of electroweak theory.
In QCD the spontaneous symmetry breakdown leads to nearly massless
Goldstone bosons (the pions) as well as the 300--400 $MeV$ of
constituent-quark mass. So in the large distance limit (momentum scales
smaller than 500--1000 $MeV$), the QCD dynamics is best described by an
effective chiral Lagrangian containing the $\sigma$, $\pi$, and
constituent-quark degrees of freedom (plus some glue) rather than the
partonic quark-gluon degrees of freedom which form the basis of
perturbative-QCD phenomenology.\cite{ref:j}

It is an extremely basic question to relate this long-distance chiral
description to the short-distance Lagrangian. The boundary between
large and short distances needs to be sharpened and quantified. And the
connection of this chiral-symmetry breaking phenomenon to confinement
needs elucidation. So far the main clue comes from the lattice: the
chiral phase transition and deconfinement phase transition in
finite-temperature QCD are indistinguishable so far.

I have devoted the final section of this talk to a description of a
specific attack on the above questions by Diakonov and his co-workers.
I am no expert in this topic. But their work strikes me as a promising
attack on the question at an impressively fundamental level, work which
respects a variety of fundamental principles. Right or wrong, I think
it is well worth careful attention and study.

Much closer to most of the material contained in the lectures at this
school is what goes on in QCD in the high-energy limit. Again we may
look at this in space-time.  But for high-energy collision dynamics the
important action is in the neighborhood of the light cone. Near the
past light cone, there is perturbative ``evolution"; it is here where
each incoming hadron is replaced, in parton-model ideology, by an
incoherent beam of incident partons which eventually scatter off a
similar ``beam" of partons in the other projectile. Near the future
light cone there occur perturbative branching processes which
create the multijet structure of typical QCD final states. Further into
the interior of the future light cone, things get messy because the
partons must find their way into final-state hadrons without violating
the nonperturbative demand of perfect confinement; never must a single
quark escape into an isolated final state. Finally, deep inside the
future light cone there may also be dynamics: some of us speculate that
this region contains a vacuum state with a rotated value of its order
parameter (disoriented chiral condensate) which decays into coherent
states of pions with curious properties.\cite{ref:k}  It is conceivable
that there could be other mechanisms of particle production from this
region of spacetime as well. This need not happen, but if it does it is
novel physics not contained in existing event generators.

The time scales for evolution of the final state in high energy
collisions is very large, proportional to the energies involved. The
time scale for hadronization of leading particles in a jet, in
reference frames where the nearest neighboring jet is 90$^\circ$ away
(The correct way, in fact, to define what is and is not in jets is to
do it in such frames.\cite{ref:l}), is proportional to the transverse
momentum or transverse energy of the jet. Thus there is a direct
correspondence between the configuration-space and momentum-space
description of jets: the production angles are of course the same,
while the (large) $p_t$'s and (large) jet-hadronization time-scales $T$
are in direct proportion.

Indeed since one can simultaneously describe the gross properties of
jet contents in both momentum space and space-time, it is clear that
the description must be macroscopic, quasi-classical in nature. The
vital region for phenomenology is the region of space-time where the
real observed hadrons are produced. In QCD this is typically a fractal
surface, because there can be jets within jets within jets $\ldots$.
Recall that in the absence of QCD jet phenomena, hadrons are produced
with rather uniform density in the lego plot. Since the lego plot area
is proportional to $\log s$, the multiplicity should rise with $s$ in a
similar way.  When additional jets populate the lego plot, they
increase the phase space area by an amount equal to $\log p_t$ (or
$\log T$) per jet. The jets themselves can contain additional jets in
their (extended) phase space, leading to a branching structure and
fractality by the time all jet evolution is accounted for. The hadron
multiplicity is then proportional to the total area of this extended
phase space, which thereby acquires fractal properties,\cite{ref:m} and
the multiplicity growth with $s$ becomes more rapid.

It is of course a challenge to provide a sharp description of all this.
And the situation is in fact quite good. There is the phenomenon of
``preconfinement", which is a perturbative mechanism which keeps color
and anticolor close together (most of the time) in momentum space as
the branching scale becomes soft and hadronization is
invoked.\cite{ref:n} The Monte-Carlo programs which employ QCD
branching mechanisms work well, and subtle, QCD-specific phenomena like
the ``string effect" are predicted and seen.\cite{ref:o} Nevertheless
some of the other claimed successes are consequences of phase space and
little more. And a purely perturbatively based approach cannot be
complete, because confinement is neglected and confinement is
important. For example, much is made of ``local parton-hadron duality"
which is the statement that the perturbatively computed momentum-space
densities of ``produced" soft partons matches smoothly to the
corresponding densities of produced hadrons. This principle is
reasonable almost always, especially when the densities are not small.
But now and then the phase space densities of produced partons will
fluctuate to small values, and some nonperturbative mechanism (\eg\
flux tubes) must intervene. For example the $Z$ occasionally will decay
into two pions and nothing else. Local parton-hadron duality asserts
that with comparable probability the $Z$ will evolve into two
final-state partons and nothing else, \eg\ a $q$ and $\bar{q}$; no
gluons choose to be emitted. But this is a disaster because at the
hadronization time of the quarks they are 50 to 100 fermis apart.  The
local duality should apply to space-time as well as momentum space, and
there is a clear problem with simple causality. One cannot be satisfied
with a theory of hadronization which accounts for confinement only most
of the time.

Finally there are challenges even within the perturbative sector. These
need only be mentioned here briefly, because they will get a lot of
attention in the other lectures. It turns out that despite the
phenomenon of asymptotic freedom, the interaction of partons at extreme
cms energies and a fixed small distance scale (say, of impact
parameter) is supposed to grow as a power of energy, perhaps almost
linearly with cms energy.\cite{ref:d} Thus at extremely high values of
$s/t$, the parton-parton interactions might become strong, with a
breakdown of perturbation theory and lots of diffractive phenomena.
This is focusing much-needed attention on diffractive phenomena,
especially short-distance, high-$p_t$ diffractive processes. The buzz
words are hard diffraction, soft and hard Pomerons, BFKL Pomerons, etc.
It is an exciting new field, as the proceedings of this school
exhibit.\cite{ref:d,ref:p,ref:q}
   
\section{Dilemmas}

It is the challenges facing QCD that makes its investigation so much
fun. But with the challenges come the dilemmas, which can sometimes
make the investigations frustrating. What follows is a rather random
potpourri of dilemmas:

On the experimental side, many of the greatest challenges lie in the
nonperturbative sector: low energy spectroscopy (\eg\ of glueballs) and
collision dynamics, as well as the problems of hadronization and
soft diffraction at high energies. Unfortunately these problems
nowadays have little sex appeal, and the interest in---and resources
for doing---low energy spectroscopy and soft-collision dynamics is
simply insufficient. For example a low-energy full-acceptance
spectrometer with modern capability would not be costly in comparison
with most modern detectors, and could by itself augment the
spectroscopy data base---much of which was established long ago via
bubble chamber techniques---by orders of magnitude. There is not even
an initiative anywhere for doing this. I am informed by Bill Dunwoodie
that there actually was a proposal not so long ago for a
full-acceptance spectrometer\cite{ref:r} at the proposed Canadian
facility KAON. But it did not survive the death of KAON itself and is
now abandoned. What a pity!

I also bemoan the lack of interest in full-acceptance,
large-cross-section physics at hadron-hadron collider energies. The
bemoanings are made in my Snowmass talk\cite{ref:a}  and
elsewhere,\cite{ref:s} and will not be repeated here.

Most of the challenges for theorists mentioned in the previous section
are low energy or soft phenomena which go beyond perturbation theory.
And there are not too many good options for theorists under those
circumstances. Lattice QCD is a very powerful way of going beyond
perturbation theory, but it is very difficult to apply to high-energy
collision dynamics.

A basic dilemma at higher energies is the problem of hadronization,
where as already mentioned there is a fuzzy boundary between what is
perturbative and what is not. The techniques for creating a precise
understanding are still lacking.

Finally there is the problem of QCD vacuum structure. Understanding the
QCD vacua (there are many of them) is the key to the question of
confinement and is important for the phenomenology of the effective
chiral theory valid in the low energy limit. Again the available
techniques are limited; nevertheless the problem is being attacked and
progress is being made. The next section on the recent work of Diakonov
and his co-workers is evidence of some of the best of this work and,
right or wrong, is an exemplar of the kind of thing that is sorely
needed to really make major new inroads into the full understanding of
QCD.

\section{Diakonov \etal: Instantons and their\hfill\break
         Consequences}

The starting point of Diakonov's work\cite{ref:au} was to study the
influence of instantons on low-energy QCD. The instanton, something not
easy to explain even at length,\cite{ref:v} no less in a short summary
like this, is a classical solution of the (Euclidean) QCD field
equations, which physically is related to the mixing (via a tunneling
mechanism) of various QCD vacua which differ by gauge transformations
of nontrivial topology. All this was discovered twenty or so years
ago,\cite{ref:w} but at that time infrared divergences in the
calculations \cite{ref:x} made quantitative consequences near
impossible to attain. More recently Shuryak determined
phenomenologically the properties that the instanton ``fluid" should
have in order to be consistent with known data.\cite{ref:y} Diakonov
and Petrov\cite{ref:u} then performed variational calculations which
supported the Shuryak picture. Since then there have appeared some
lattice calculations of instanton effects which, although still
somewhat controversial, appear to lend support as well.\cite{ref:z} The
net result is that by now there is a credible picture of what the
instanton effects are. The immediate ones are the solution of the
$U(1)$ problem (why the $\eta^{\,\prime}$ meson is so heavy) and the
existence of a gluon condensate (seen in QCD sum rules, yet another
piece of QCD not discussed at this school\cite{ref:aa}).

All this, however, is still rather abstract; it is not clear what more
these abstruse considerations have to say to the experimentalists in
the trenches. However the next steps taken are more directly related to
QCD phenomenology. The most relevant features, in my opinion, are as
follows:

\begin{enumerate}
\item
The quark-parton degrees of freedom are influenced by the presence of
the instantons, and they get a ``constituent-quark" mass as a
consequence of having to propagate through the instantons.

\item
The above mechanism leads naturally to spontaneous breaking of the
strong-interaction chiral symmetry.

\item
Therefore there must be the Goldstone degrees of freedom
(almost-massless pions) in the spectrum as well as the constituent
quarks.

\item
The low-energy chiral effective Lagrangian can be constructed.  The
lowest order terms are universal (model independent) in form, depending
only on symmetry considerations. However, higher order terms are also
present and can be estimated. The magnitudes of these terms are in
agreement with what is needed for the phenomenology.

\item
The Goldstone pions can be shown to be composites of the constituent
quarks. This is an improvement on the scheme put forward by Manohar and
Georgi\cite{ref:bb} some time ago. They argued for the chiral
constituent-quark-plus-pion picture based on the success of the
additive quark model. They had, however, an awkward time in
understanding whether their Goldstone pion is the same as, or distinct
from, the $^1S_1$ partner of the $^3S_1$ $\rho$. In the
Diakonov instanton picture they are not distinct.

\item
Diakonov \etal\  in addition put forward an interesting model of
baryons, which is a variant of the somewhat popular Skyrmion picture,
and to my eye an improvement.\cite{ref:cc} They assume that the pion
cloud surrounding the three constituent quarks of the baryon has a
nontrivial ``hedgehog" topology, as originally suggested by Skyrme long
long ago\cite{ref:dd} (see Eq. (\ref{eq:K}) below). Then it can be shown
that in such an external field there will be one and only one quark
bound state with energy in the gap between the continua starting at $E
= + m$ and $E = - m$. This state can be populated with one quark of
each color to make objects with the quantum numbers of the nucleons. In
the large $N_c$ limit the combined wave functions of these quarks can
be treated {\it a la} Thomas-Fermi atomic theory as a source of the
``hedgehog" pion field, leading to a self-consistent semiclassical
description of the nucleon. To recover quantum mechanics, in particular
the classification of the energy levels, the ``cranking-model"
techniques of nuclear theory can be employed to give a reasonable
description.\cite{ref:ee} So this picture has a quite good formal
justification in the large $N_{c}$ limit.

\item
Finally, with this picture of the nucleon, they calculate\cite{ref:ff}
the distributions of the ``primordial" partons within the nucleon,
namely the leading twist parton distributions at a low value of $Q^2
\approx 0.5\ GeV^2$, which when evolved to higher $Q^2$ via the DGLAP
evolution equations give the leading twist contributions to the
structure functions. Their results agree reasonably with the Gluck,
Reya, Vogt\cite{ref:gg} primordial parton distribution functions which
are input by hand in order to reproduce deep-inelastic scattering data.
But more important in my opinion is the way Diakonov \etal\ can
maintain the internal consistency of the formalism. The validity of a
variety of current algebra sum rules is established. This is highly
nontrivial, because relativistic effects are very important, and
valence antiquark distributions must be present; they are created by
the back reaction of the nucleon valence quarks on the pionic
``hedgehog" sea. The techniques which are employed provide valuable
lessons for all bag-model descriptions of hadrons.

\end{enumerate}

There remains a major missing link: an understanding of confinement.
The effects of gluons in this low-energy limit are formally of ``higher
order". In one sense this is good, because in the low energy
constituent-quark spectroscopic world the gluon degrees of freedom do
not seem to play a central role. On the other hand their effects cannot
be omitted, because confinement depends on them. Probably some of the
effective quark mass is accounted by something akin to color flux-tube
energy, and the dynamical effects of fluctuations about an average
value are not too important. I think the ideal arena for studying this
problem is that of heavy-flavor mesons and baryons, where the source of
color is static and understood (a stationary heavy $b$-quark, or Wilson
line) and only how the color finds its way into the single constituent
quark degree of freedom of the $B$-meson, or alternatively into the
``Skyrmionic" quark-baryon wave function, needs to be solved. Some work
has been done, but more is needed.\cite{ref:hh}

While there may well be reason to exhibit skepticism regarding the
whole Diakonov program, I still want to emphasize that this kind of
work is at the most important forefront of QCD. It links the confining
world to the perturbative sector. Most of the known nonperturbative QCD
phenomena are involved, and the work touches upon the edge of some of
the best perturbative phenomenology which exists, namely the information
on deep-inelastic structure functions. The level of attack is much
deeper than mere phenomenological model building. It deserves, I
believe, close attention and constructive criticism.

\section{Some more details on the Diakonov program}

The preceding description was very general in nature, and what follows
is a slightly more technical version of some of the same material. It
is far from definitive, if for no other reason than the limited
competence of yours truly. However there are recent lecture notes
\cite{ref:au} to consult for a more detailed and authoritative
version.

\subsection{What about these instantons?} 

As already mentioned, an instanton is a solution of the QCD classical
field equations in Euclidean space-time with finite action. It contains
a ``topological knot" and is localized in space-time. It also has a
size parameter which can take any value in principle. The immediate
function of these instantons is to create couplings, via tunneling,
between different Minkowski-space QCD vacua, vacua which differ from
each other by a gauge transformation which also contains a
``topological knot". Because of these nonperturbative
tunneling couplings, the many initially degenerate QCD vacua, which
can be classified in terms of the number of gauge knots they contain,
are coupled together and must be diagonalized, leading to the so-called
$\theta$-vacua, which are the true energy eigenfunctions of the vacuous
QCD theory.

When theorists initially attempted to estimate the magnitude of these
effects they were thwarted by the presence of large numbers of large
instantons, whose effects were not under control. Shuryak, working
phenomenologically, argued that if instantons with sizes larger than
about 0.3 fermis (or a momentum scale $\approx 600\ MeV$) were
suppressed, instanton-induced phenomenology could be understood.
Furthermore, were this true, the instanton ``liquid" in Euclidean
space-time would be dilute, in the sense that the mean separation $R$
between instantons would be 2 to 3 times larger than the important
instanton size $\rho$. As already mentioned, Diakonov and
Petrov,\cite{ref:u}  using variational techniques, found a candidate
mechanism for this to happen, namely medium-range
instanton-antiinstanton repulsion.

The bottom line is that the effects of large instantons are arguably
damped out at a known scale, with a bonus of a small parameter (the
instanton packing fraction in Euclidean space-time) in the formalism.
This then becomes the working hypothesis for going further. It is not
rigorously established but is credible.

\subsection{How do the instantons induce chiral symmetry 
\hfill\break breaking?}

The next step is to introduce the quarks and calculate their influence.
The equation of motion of quarks in a classical instanton field (again
in Euclidean space time) also shows a remarkable feature---the
existence of ``zero-mode" solutions of the Dirac equation of the
quark in the presence of the instanton (with zero eigenvalue of the
Euclidean Dirac operator) and which are localized around the instanton.
Just as for the instanton itself, the implication of these solutions
for physics is subtle and deep. For example they influence the presence
(or absence) of $CP$ violation in the strong interactions. The vital
buzzword here is ``spectral flow": a filled negative-energy level (now
in Minkowski space) in the negative energy sea can, because of the
knotty gauge potentials, be pushed above zero (in the chiral limit of
massless quarks), while other empty positive energy levels with
different quantum numbers can be pushed into the negative energy
sea.\cite{ref:ii} The net result is that there can be net pair-creation
induced, with the pair not necessarily having vacuum quantum numbers.
All this activity is quite sufficient to create the mechanism of
spontaneous symmetry breakdown.

In the calculations which argue for spontaneous symmetry breaking, it
is necessary to include the mixings of zero modes associated with
different instantons, something rather nontrivial. What follows are a
few equations for theorists and well-educated experimentalists to give
a flavor of what is done. The information about all this kind of thing
is to be extracted from the Euclidean partition function
\begin{eqnarray}
Z &\sim& \int \D A\, e^{-\frac{1}{g^2}\int d^4x\, F^2}\int \D
\psi\calDslash\bar \psi e^{\int d^4x\bar
\psi(\,\not{\hbox{$\scriptsize\nabla$}}-g\ \Aslash-m)\psi}\nonumber \\
&\equiv& \VEV{\det (\nablaslash-g\Aslash-m)} 
\Rightarrow \VEV{e^{\frac{1}{2}\sum_n \ell n
(\lambda^2_n+m^2)}} 
\label{eq:A}
\end{eqnarray}
where in the second line the Gaussian integral over fermionic quark
fields is performed, and where
\begin{equation}
\lambda_n = n_{th}\ \mbox{eigenvalue of} \ (\nablaslash-g\Aslash)
\label{eq:B}
\end{equation}
and $m$ the small quark-parton mass of a few $MeV$.

There is one zero eigenvalue per instanton per quark flavor in the
dilute-instanton approximation. But when the effect of the overlapping
of zero modes from separate instantons is taken into account, the zero
eigenvalues repel. The typical values become
\begin{equation}
\VEV\lambda \sim \frac{\rho^2}{R^3} \ .
\label{eq:C}
\end{equation}
Now by definition the chiral order parameter is
\begin{eqnarray}
V\VEV{\bar \psi\psi} &\simeq& \frac{\partial}{\partial m}\, \ell n\, 
Z\ \Big|_{m=0} \nonumber \\[1em]
&=& \frac{\partial}{\partial m}\, \VEV{\frac{1}{2}\int d\lambda\,
\nu(\lambda)\, \ell n\, (\lambda^2+m^2)} \nonumber \\[1em]
&=& \VEV{\int d\lambda\,
\nu(\lambda)\left(\frac{m}{\lambda^2+m^2}\right)}_{m\rightarrow 0}
\label{eq:D}
\end{eqnarray}
where we go to continuum normalization via
\begin{equation}
\sum_n \Leftrightarrow \int d\lambda\, \nu(\lambda) 
\label{eq:E}
\end{equation}
and $V$ is the (Euclidean) space-time volume.

This shows that chiral symmetry breaking will occur provided the
density of zero modes $\nu(\lambda)$ at $\lambda=0$ is nonvanishing.
But this is what is estimated to occur,
\begin{equation}
\VEV{\bar\psi\psi} \sim \frac{\nu(0)}{V} = \frac{N}{\VEV\lambda V}\sim
\left(\frac{R^2}{\rho}\right)\, \left(\frac{1}{R}\right)^4 =
\frac{1}{\rho R^2} \ .
\label{eq:Ea}
\end{equation}
The factor $\rho/R^2$ for $\VEV\lambda$ occurs because
\begin{equation}
\VEV\lambda^2 \Rightarrow \VEV{\lambda^2} = N\VEV{\lambda^2}_0 = N \int
\frac{d^4R}{V}\, \left(\frac{\rho^2}{R^3}\right)^2 = \frac{N\rho^2}{V} =
\frac{\rho^2}{R^4} \ 
\label{eq:F}
\end{equation}
with $\VEV{\lambda^2}_0$ the contribution to $\lambda^2$ from one
instanton. Note that  this happens because we have added the
contributions to splittings from all the neighboring instantons in
quadrature. This rough argument, due to Diakonov, actually can be
refined, so that the conclusion is quite robust.

The parameters of the constituent quarks can be estimated from the 
instanton parameters, which are
\begin{equation}
\mbox{instanton size:}\qquad \bar\rho \approx 0.3\, f
\label{eq:G1}
\end{equation}
\begin{equation}
\mbox{instanton spacing:}\qquad R \approx 1\, f \ .
\label{eq:G2}
\end{equation}
This means that the fraction of Euclidean space-time occupied by
instantons is
\begin{equation}
 \pi^2 \left(\frac{\rho}{R}\right)^4 \approx 0.1 \ .
\label{eq:G3}
\end{equation}
The quark mass, in order of magnitude, turns out to be
\begin{equation}
M_Q \sim \frac{\rho}{R^2} = \frac{1}{\rho}\,
\left(\frac{\rho}{R}\right)^2
\label{eq:H}
\end{equation}
and the careful calculations produce a reasonable value of constituent
quark mass of 350--400 $MeV$.

The pion decay constant $F_\pi$ can also be estimated 
\begin{equation}
F_\pi \cong \frac{{\rm const}}{\rho} \cdot
\left(\frac{\rho}{R}\right)^2 \sqrt{\ell n\, \frac{R}{\rho}}
\approx 100\ MeV \ .
\label{eq:I}
\end{equation}
In the very low momentum limit, the constituent quark degrees of
freedom can be integrated out of the partition function, leaving a
chiral effective action of the form
\begin{equation}
Z = e^{iS(\pi)} = \int \D\psi\D\bar\psi\, \exp
i\int d^4x \left(\bar\psi_L \left[i\nablaslash-M\,
e^{i\vec\tau\cdot\vec\pi}\right]\, \psi_R \ + {\rm h.c.}\right) \ .
\label{eq:J}
\end{equation}
There is also a ``gap equation" relating how the pionic degrees of
freedom are related to the quarks, but I have had difficulty dredging
the details out of the easily available literature.

\subsection{How does this lead to a model of the nucleon?}

Thus far it has been sufficient to look at the theory in Euclidean
space-time, a clear indicator that phenomenology is somewhat distant.
The reason for the success is that the theory has been about the vacuum
properties much more than about excitations of the vacuum, where
Minkowski-space description is essential. (If the energy of the system
is zero, then its analytic continuation to imaginary energies does not
change too many things.) Nevertheless the Euclidean analysis has led to
an effective action, which can be continued to Minkowski space-time and
used for dynamics.

The model of the nucleon is built from this action via the Skyrme
ansatz for the pion ``condensate":
\begin{equation}
U \equiv e^{i\vec\tau\cdot\vec\pi} = e^{i\vec\tau\cdot\hat rf(r)}
\label{eq:K}
\end{equation}
with
\begin{equation}
f(0) = \pi \qquad f(\infty) = 0 \ . 
\label{eq:L}
\end{equation}
Because 
\begin{equation}
U(0) = - 1
\label{eq:M}
\end{equation}
and
\begin{equation}
U(\infty) = + 1
\label{eq:N}
\end{equation}
the pion field contains the ``topological knot"; $U$ cannot be 
continuously deformed to the unit matrix. 

Now the Dirac equation is solved in this pion field
\begin{equation}
i\nablaslash - MU(r) = 0
\label{eq:O}
\end{equation}
and, as already advertised, one bound state is found to exist with
$|E|<M$. The bound-state wave function is then determined by
calculating the summed energy of the negative-energy Dirac sea and the
bound state contribution as a function of the trial function $f(r)$,
and then minimizing with respect to the choice of $f$. The resulting
structure is classical, and the quantum structure is built by using the
``cranking model", \ie\ projecting the constructions on eigenfunctions
of rotations and translations. The nucleon and $\Delta$ masses can be
calculated; the nucleon mass is somewhat on the high side (1200 $MeV$
or so), although there are several candidate apologies for this
situation. With this model, a variety of nucleon static properties are
calculated with reasonable success.

\subsection{What implications does this have for deep-inelastic
structure functions?}

An especially interesting application of the model is in the
construction of the primordial parton distributions, 
defined as follows:\cite{ref:jj}
\begin{equation}
\left.\begin{array}{rl} q(x) & x>0 \\ -\bar q(-x) & x<0
\end{array}\right\} = \frac{1}{4\pi}\int^\infty_{-\infty} dt\,
e^{ixMt}\VEV{P\,|\,\psi^\dagger(0)(1+\gamma^0\gamma^2)\, 
\psi(y)\,|\,P}
\ .
\label{eq:P}
\end{equation}
with
\begin{equation}
y = (t,-t,0,0) \ .
\label{eq:Pa}
\end{equation}
This is to be interpreted as the input parton distributions at the
highest value of scale allowed by the effective chiral theory, namely
the scale associated with the typical instanton size, 600 $MeV$, or
$Q^2 \approx 0.4\ GeV^2$. Note that it is defined in the nucleon rest
frame, but when boosted to an infinite-momentum frame becomes the usual
correlation function defining the parton distributions.

\begin{figure}[htb]
\begin{center}
\leavevmode
{\epsfxsize=4in \epsfbox{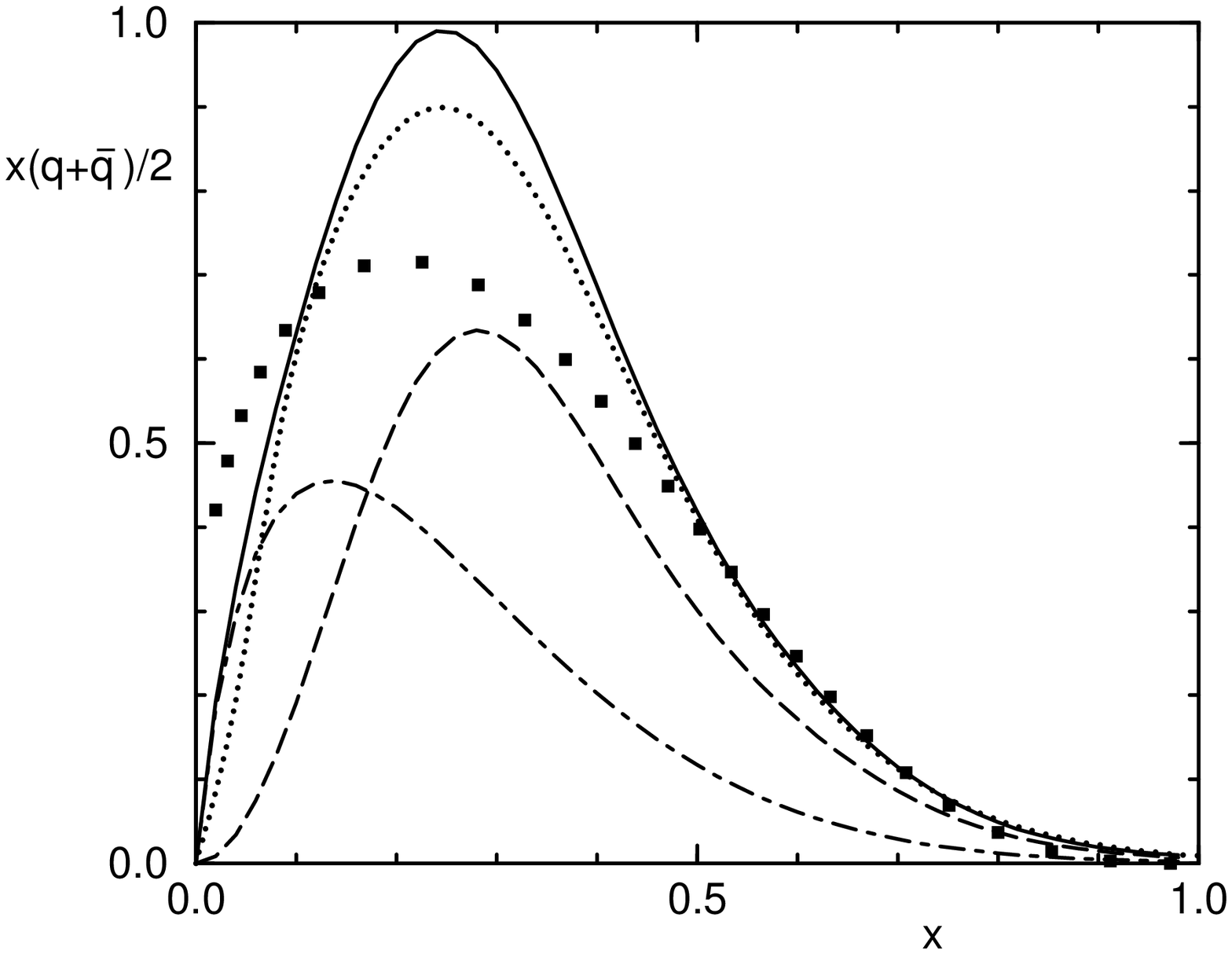}}
\end{center}
\caption[*]{The singlet unpolarized distribution, $x[u(x)+d(x)+\bar
u(x)+\bar d(x)]/2$. Dashed line: regularized contribution from the
discrete level; dash-dotted line: contribution from the Dirac
continuum; solid line: the total distribution, namely the sum of the
dashed and dash-dotted curves, dotted line: the exact total
distribution; squares: the parametrization of Ref.~32.} 
\label{fig1}
\end{figure}

\begin{figure}[htb]
\begin{center}
\leavevmode
{\epsfxsize=4in\epsfbox{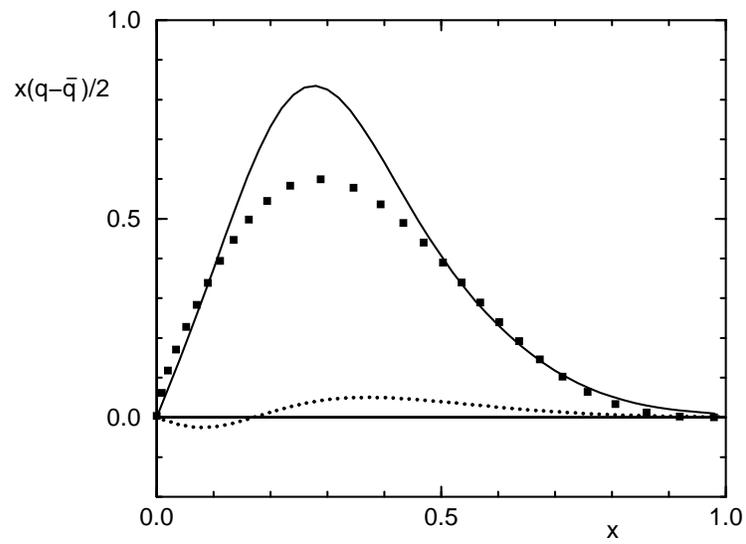}}
\end{center}
\caption[*]{The baryon number distribution, $x[u(x)+ d(x) -\bar
u(x)-\bar d(x)]/2$. Solid line:  distribution from the unregularized
discrete level; dotted line: exact Dirac continuum contribution;
squares: the parametrization of Ref.~32.} 
\label{fig2} 
\end{figure}

\begin{figure}[htb]
\begin{center}
\leavevmode
{\epsfxsize=4in\epsfbox{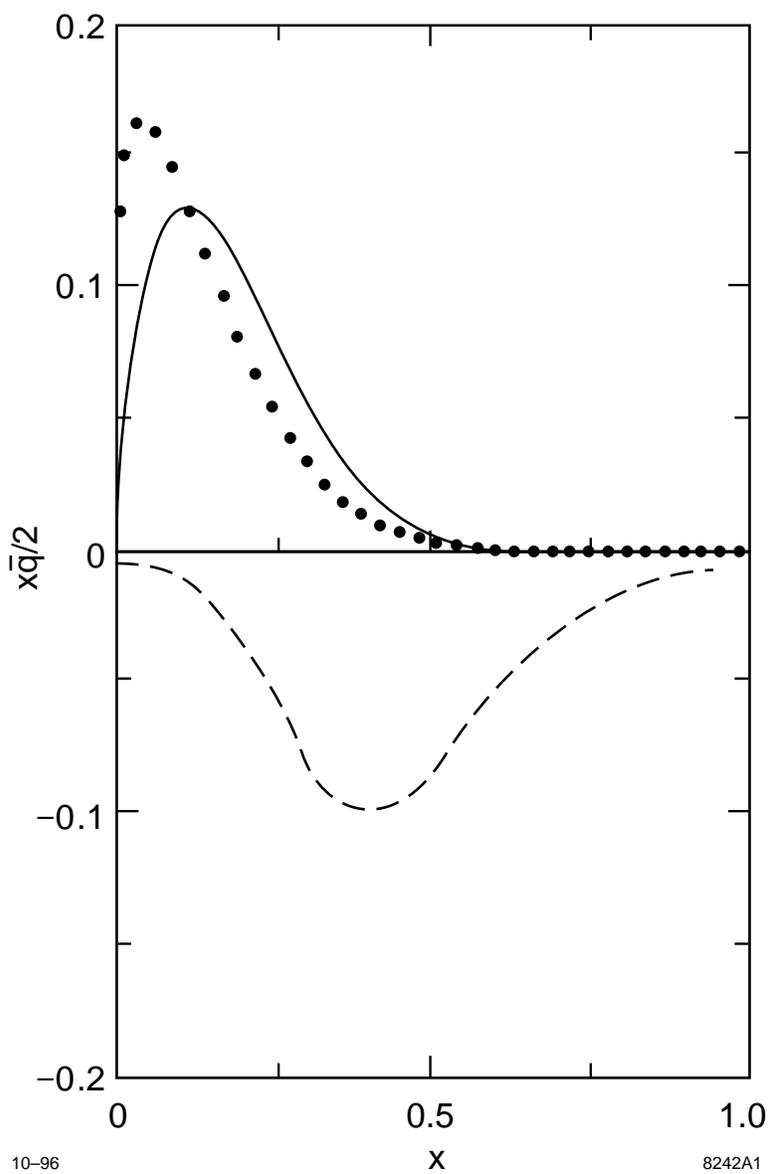}}
\end{center}
\caption[*]{The antiquark distribution, $x[\bar u(x)+\bar d(x)]/2$. Solid
line: theory; squares: the parametrization of
Ref.~32; dashed line, contribution from the discrete level only.} 
\label{fig3} 
\end{figure}

\begin{figure}[htb]
\begin{center}
\leavevmode
{\epsfxsize=4in\epsfbox{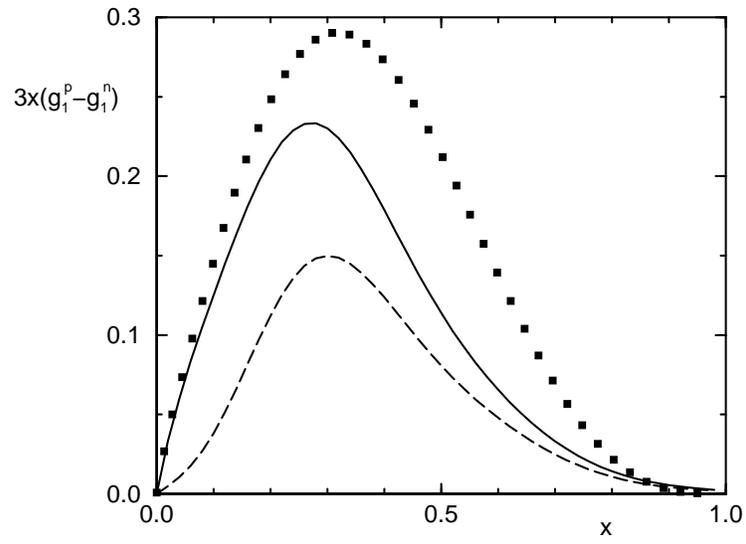}}
\end{center}
\caption[*]{The isovector polarized distribution, $x[\Delta u(x)-\Delta
d(x)+\Delta\bar u(x)-\Delta\bar d(x)]/2$. Dashed line: regularized
contribution from the discrete level; solid line: the sum of the
contributions from the discrete level and from the continuum; squares: 
the parametrization of Ref.~32; dashed line: contribution from the
discrete level only.} 
\label{fig4} 
\end{figure}

\begin{figure}[htb]
\begin{center}
\leavevmode
{\epsfxsize=4in\epsfbox{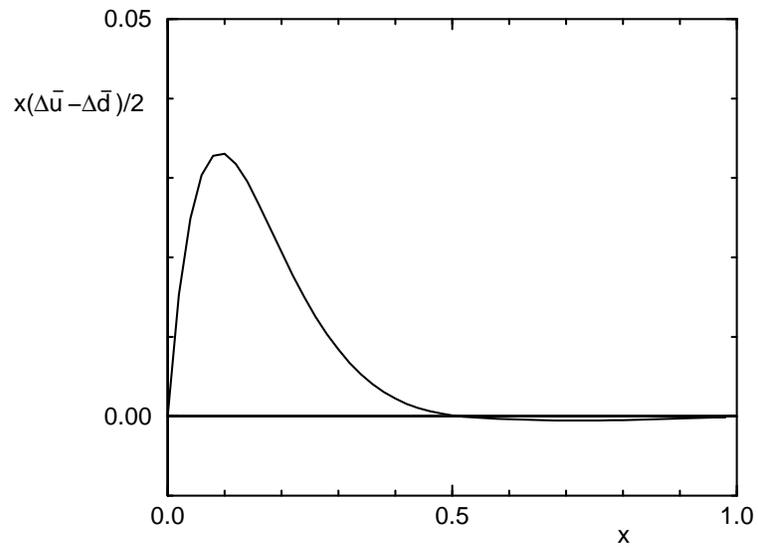}}
\end{center}
\caption[*]{The isovector polarized distribution of antiquarks,
$x[\Delta\bar u(x)-\Delta\bar d(x)]/2$. Reference~32 assumes
this quantity to be zero.} 
\label{fig5} 
\end{figure}

Note that the definition in Eq. (\ref{eq:P}) admits the introduction by
necessity of valence antiquark distributions. And, as mentioned
earlier, the contribution of the discrete level by itself leads to
negative-definite valence antiquark distributions. It is necessary to
calculate the (distorted) negative-energy continuum contributions
before obtaining sensible results. When this is carefully done, the
antiquark distributions happily are positive definite. Some of these
are shown in Figs. \ref{fig1}--\ref{fig5}. In particular, in Fig.
\ref{fig3}, which exhibits the flavor singlet antiquark distributions,
is sketched the negative contribution of the discrete level, as well as
the summed result.

A variety of deep-inelastic sum-rules are also tested, and shown to be
in principle (as well as numerically) satisfied. These include the sum
rules for baryon number, momentum (at this level all momentum is
carried by quarks), isospin, and flavor-nonsinglet polarized
distributions. Also, the Gottfried sum, which measures the flavor
non-singlet antiquark distribution, is calculated and has nonvanishing
right-hand side, with the sign needed to account for the data. The
argumentation for these results goes deep into the basic structure of
the model, and the consistency is very satisfying.

It would be a great advance if the description of mesons, for which
there is no Skyrmionic topological starting point, could be carried to
the same level of sophistication. Are mesons really so different from
baryons? I think the best candidate for study is the $B$-meson. If
progress can be made there, it may also shed light on the confinement
issue, which so far has remained beyond the scope of these methods.

\vspace{.5cm}

\vspace{.5cm}

\vspace{.5cm}

\vspace{.5cm}

\vspace{.5cm}

\end{document}